# Beyond Technological Usability: Exploratory Factor Analysis of the Comprehensive Assessment of Usability Scale for Learning Technologies (CAUSLT)

*Jie Lu[1]

Matthew Schmidt[1]

Jinnie Shin[2]

[1]Department of Workforce Education and Instructional Technology, University of Georgia

[2]School of Human Development and Organizational Studies in Education, University of Florida

*Corresponding author: Dr. Jie Lu, jlu@uga.edu

**Abstract**

Traditionally rooted in the domain of Human-Computer Interaction (HCI), usability has been primarily associated with the technological performance of a system's user interface. However, as learning technologies continue to advance, a pressing need exists to evaluate these tools from a broader perspective, encompassing not just technological but also pedagogical and sociocultural dimensions. The current paper delves into the multifaceted nature of usability in the context of Learning Design and Technology (LDT). We identified prevailing gaps in current usability research practices within LDT, notably the over-reliance on HCI-derived instruments that may not holistically capture the unique usability demands of learning technologies. To address these challenges, we embarked on the development and analysis of the Comprehensive Assessment of Usability Scale for Learning Technologies (CAUSLT). A total of 155 responses were collected and analyzed. Utilizing exploratory factor analysis, this study aimed to explore core constructs for the development of CAUSLT. Our findings underscore the importance and the critical need for a comprehensive usability evaluation framework tailored for learning technologies, setting the stage for more effective and user-centric educational tools.

*Keywords*: Usability research, human-computer interaction, exploratory factor analysis, instrument development

## 1. Introduction

Usability is a pertinent concept commonly associated with software engineering (Bender et al., 2021; Cayola & Macías, 2018; Curcio et al., 2019) and technology development (Holthe et al., 2018; Maramba et al., 2019; Moon et al., 2019). Originating in the field of Human-Computer Interaction (HCI), usability has been conceived of as a measure of the performance of a given system user interface (UI). A series of standards (e.g., ISO 25010, ISO 9241) have been established to define and characterize usability for this purpose. As defined by the International Standards Organization (ISO), usability pertains to the degree to which a system can effectively and efficiently assist users to accomplish predetermined goals with satisfaction (ISO, 2011). Regarding the attributes of quality in use of a product or system (ISO, 2018), five characteristics have been identified: (1) effectiveness, (2) efficiency, (3) satisfaction, (4) freedom from risk, and (5) context coverage. Consequently, a plethora of sophisticated usability evaluation methods and instruments have been developed to quantify usability, including the System Usability Scale (SUS; Brooke, 1996), the Computer System Usability Questionnaire (CSUQ; Sauro & Lewis, 2016), and the Software Usability Measurement Inventory (SUMI; Kirakowski, 1996).

As learning technologies continuously evolve to advance learning across a variety of fields, an increasing need has emerged to evaluate such tools not only from the perspective of ease-of-use, but also from the perspective of individual users' experiences while they engage in technology-supported learning. The importance of conducting usability evaluations for learning technologies is well-established within the field of Learning Design and Technology (LDT) (Reeves et al., 2002; Reeves & Hedberg, 2003; Squires & Preece, 1996, 1999). However, the current practice of usability research in this field is problematic in two ways. Firstly, usability instruments are primarily borrowed from the field of HCI (e.g., SUS, CSUQ). Although these instruments adeptly measure the technological aspect of usability, they do not account for learning-specific features (i.e., pedagogical and sociocultural aspects of usability) (Lu et al., 2022). Researchers and practitioners in the field of LDT have criticized the narrow focus on the technological aspect of usability when evaluating learning technologies (Jahnke et al., 2020; Moore et al., 2014; Nokelainen, 2005; Pham et al., 2019; Tervakari & Pohjolainen, 2013). This criticism stems from the recognition that, within a learning context, other aspects of usability are equally important in fostering effective, efficient, and satisfying learning experiences. These aspects encompass learning designs (Zaharias, 2004; Zaharias & Poylymenakou, 2009), social interactions (Heldal, 2007; Preece, 2001), cultural considerations (Amant, 2017; Reeves, 1994), and the overall user experience (Devy et al., 2017; Quiñones et al., 2018). Well-established and widely-used instruments like the CSUQ and SUS primarily emphasize the technological aspects, providing only a narrow perspective on the overall usability of such learning tools. However, educational and learning technologies require broader considerations (Lu et al., 2022). Indeed, a system with remarkably high technological usability does not necessarily translate into an effective, efficient, or satisfying learning experience that leads to desired learning goals. Take the use of virtual reality (VR) systems in clinical settings as an example (Ayoub & Pulijala, 2019; Pottle, 2019), the highly sophisticated system may feature incredibly realistic simulations, allowing students to interact with virtual atoms and particles in an intuitive way. However, this does not necessarily lead to effective learning (Mayer et al., 2023). The focus on high-definition simulations could overshadow the core learning objectives, with learners spending more time on the virtual world rather than focusing on the concepts they need to master. Consequently, while the VR system might boast high technological usability with a state-of-the-art interface and features, it does not automatically translate to an effective, efficient, or satisfying learning experience.

The second concern regarding current practice in usability research centers around the psychometrics of instrument development. Numerous usability studies reported in the field of LDT applied either self-developed instruments or modified versions of existing instruments (Lu et al., 2022; Lu, 2025), however, attempts to develop new usability instruments that more accurately capture the characteristics of learning technologies have generally failed to follow a formal and rigorous procedure, resulting in concerns regarding the validity and reliability of these measures (Quiñones & Rusu, 2017). Specifically, researchers in the field of LDT often create their own instruments without adequate evaluation, leading to uncertainties about the quality of these tools and raising concerns regarding the accuracy and consistency of results.

Therefore, there is a need for instrumentation that is methodologically sound, valid, and reliable in quantifying the multifaceted nature of usability for learning technologies. Such an instrument should be attuned to the complexities inherent in evaluating the usability of learning technologies, extending beyond the purely technological aspect of usability to encompass pedagogical and sociocultural aspects of learning technology usability. To address these issues, an iterative process of instrument development was followed to systematically design, develop, implement, and evaluate a survey instrument that is tailored for assessing multiple aspects of the usability of learning technologies. This study was guided by the following research question:

Given the aim to capture the multifaceted nature of learning technology usability, what is the underlying factor structure of the developed usability survey instrument?

## 2. Literature Review

### 2.1 Usability Research in the field of LDT

To date, the most cited definition of usability is the ISO standard, which includes usability studies conducted within the field of LDT (Lu, 2025). Nevertheless, some researchers have already sought to ground the notion of usability in the contexts of learning. An early attempt and precedent work can be found in Reeve's 14 Pedagogical Dimensions (1994), where he proposed 14 dimensions of pedagogical considerations, covering a range of factors that are relevant to the effectiveness and quality of computer-based education (CBE), as evaluation criteria under the context of learning. These dimensions provide an actionable framework for assessing the usability of CBE in the context of learning, where usability considerations are intertwined with each dimension, ensuring that the design and functionality of the CBE support effective learning experiences and align with the pedagogical principles and goals of the learning program.

Reeves and colleagues (Reeves et al., 2002) introduced an instrument along with a protocol specifically designed for instructional designers engaged in heuristic evaluation of e-learning programs. This effort involved modification and expansion of the original ten heuristics for general software, developed by Nielsen (1995), to create new heuristics that were more specific to the evaluation of e-learning programs. In particular, a small group of experts (typically ranging from three to five) was invited to assume the role of learners and interact with the program. Their task was to assess the appropriateness in terms of UI and other aspects of usability, including learning design, instructional assessment, and resources, among others. This application of usability evaluation enables a systematic examination of e-learning programs based on established principles, aiming to identify potential and existing usability issues and improve the overall quality of the e-learning program. Specifically, feedback and evaluation outcomes can be leveraged by instructional and learning designers, allowing them to make informed decisions and enhance the usability of the e-learning program.

Building upon Nielsen and Molich's (1990) conceptualization of usability and Reeves' (1994) work on pedagogical dimensions, Nokelainen (2006) further explored the category of usefulness as integrated usability, referring specifically to Nielsen's construct of utility as pedagogical usability and his construct of usability as technical usability. In the context of pedagogical usability, Nokelainen developed a model consisting of ten pedagogical dimensions that contribute to the overall usefulness of a learning technology. These dimensions capture various aspects of the learning experience and highlight the importance of pedagogical considerations in assessing the usability of learning technologies.

## 2.2. Usability Constructs Measured in the Field of LDT

Although the criticism of evaluating only the technological aspect of usability in the field of LDT has been present for a long time, empirical research investigating the constructs intended in existing usability instruments employed in learning technology research is limited. In the context of learning technologies, findings from a systematic literature review (Lu et al., 2022) indicate that intended by the commonly used usability methods (i.e., heuristic evaluation, task analysis), the perceived usability seems to be the most prevalent usability construct. For example, Nielsen's Usability Heuristics (1995) is a widely accepted inspection method for evaluating the perceived usability of a given UI. Other frequently used standardized measures such as the SUS (Brooke, 1996) and the CSUQ (Sauro & Lewis, 2016), although evaluating constructs like learnability and interface quality, still focus on the perceived usability of the software in general. Satisfaction is the second most frequently assessed construct in usability studies conducted on learning technologies. Following this are efficiency, operability, learnability, and aesthetics, which all are usability metrics defined in the ISO standard. Learning-related attributes such as content and assessment are far less commonly evaluated usability constructs. This has led to calls to extend the usability evaluation of educational technologies to more intentionally include pedagogical considerations (Granić & Ćukušić, 2011; Jahnke et al., 2020; Pham et al., 2019), as well as social interactions (Heldal, 2007; Preece, 2001) and the social roles of users (Gamage et al., 2020). This has signaled tensions about using instruments developed outside of the field of LDT to evaluate learning technologies, which is discussed below.

**2.3 Usability Evaluation Methods in the Field of LDT**

Usability evaluation has received substantial attention as a multidimensional methodology (Moradian et al., 2018) in the field of HCI (Issa & Isaias, 2015). Consequently, methods and processes of usability evaluation developed in HCI have evolved to become relatively sophisticated (c.f., Ivory & Hearst, 2001). In an extensive survey conducted on usability evaluation methods by Ivory and Hearst (2001), five usability method classes were established, with each class containing a range of specific methods. Of the classified method classes, (1) inquiry (e.g., questionnaires), (2) inspection (e.g., heuristic evaluations), and (3) testing (e.g., task-based user testing) are the most commonly used method types in the field of LDT with the most prevalent method type being "questionnaires" (Lu et al., 2022). This is because of the efficiency and cost-effectiveness of using questionnaires to collect perceptual information from the target audience.

**2.4 Towards a Multifaceted Usability Framework for Learning Technologies**

The use of usability methods and instruments developed from outside of our field to evaluate learning technologies has introduced tensions about using instruments developed outside of the field of LDT (Tselios et al., 2008; Zaharias & Poylymenakou, 2009). Specifically, applying these methods and instruments ignores or overlooks fundamental differences between general information technologies and learning technologies (Lu et al., 2022). Responding to this tension, Jahnke and colleagues (2020) analyzed 13 articles with usability evaluation criteria and categorized them into three aspects: (1) the technological aspect (e.g., Nielsen's 10 heuristics, navigation, learnability), (2) the pedagogical aspect (e.g., learner context, learning activities, learner control, tasks), and (3) the social aspect (e.g., collaboration, communication, roles and/or relationships, social interactivity). Consequently, a three-dimensional, interconnected sociotechnical-pedagogical (STP) framework for learning experience design was proposed (Jahnke et al., 2020). This framework was later extended by Schmidt et al. (2022) to consider not only the role that the social aspect of usability plays in usability evaluation but also the role that sociocultural aspects play.

**Figure 1.**

*Interconnected system of sociotechnical-pedagogical usability (Schmidt et al., 2024)*

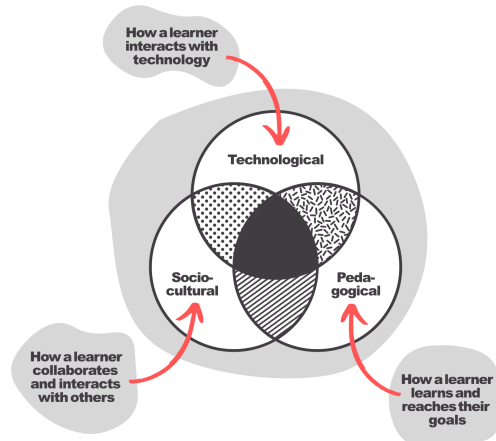

A group of researchers have put forth arguments that highlight the inadequacy of relying solely on technological usability assessment for evaluating learning technologies (Nokelainen, 2006; Squires & Preece, 1996; Tselios et al., 2008). These arguments stem from previous appeals for incorporating pedagogical usability within usability evaluation literature (Costabile et al., 2005; Nokelainen, 2006; Pham et al., 2019; Reeves, 1994; Silius et al., 2003; Zaharias & Poylymenakou, 2009) and an increased awareness of the significance of sociocultural factors, emotions, and the holistic user experience in the broader learning context (Marell-Olsson & Jahnke, 2019; Moore et al., 2014; Nakamura et al., 2018; Quiñones et al., 2018). The scope of technological usability research is constrained by its generalized applicability to users of various products or systems. However, scholars propose that when evaluating learning technologies, the focus should be on a specific user group (i.e., learners), engaged in learning-related tasks, and utilizing a particular kind of technology (i.e., educational technology; Schmidt et al., 2024). Consequently, relying solely on technological usability assessment might not adequately encompass the intricate, multifaceted, and interconnected interactions of learners as they engage with both the learning environment (the technology) and the learning space.

**2.5 Culture influences usability**

Researchers (Ford et al., 2005) looked into the role of culture in the field of HCI and argued that culture influences the way people interact with computers. For instance, culture shapes an individual's preferences for both communication and behaviors. As the interaction between people and computer systems requires communications typically as input and output, users' preferences and behaviors of communication are primarily predetermined by culture which is formed from living in a specific social environment (Ford & Kotzé, 2005). In the literature, several cultural theorists proposed dimensions of culture (Hall, 1973; Hofstede, 2011; Smith et al., 1996), with Hofstede's five dimensions (2011) being most frequently associated with usability (Downey et al., 2005; Ford & Kotzé, 2005). The dimensions are, (1) power distance, (2) uncertainty avoidance, (3) individualism-collectivism, (4) masculinity-femininity, and (5) time orientation. In his view, culture refers to patterns in which a particular group of people were programmed to think, feel, and act. Accordingly, culture differs in symbols, rituals, values, etc. Drawing from Hofstede's five dimensions, cultural aspects are clearly impacting user experience, including interaction, appearance, navigation, and more, which directly influences how users perceive usability. Consequently, interactive programs and systems must accommodate and support culturally diverse users.

Cultural usability, which emerged in the 2000s (Clemmensen, 2009; Lachner et al., 2015; Vatrapu, 2008; Zaharias, 2008), is not a new concept, but has received remarkably little attention both in the academic literature and in practice. Previous attempts to integrate cultural-related usability into evaluation (Diaz et al., 2013; Downey et al., 2005; Lachner et al., 2015; Shin, 2012; Zaharias, 2008) have contributed to raising awareness in usability research regarding the role culture plays in UX. This finding from the literature is also consistent with Gray's (2020) suggestion to adopt design strategies that better reflect the real-life experience, namely, social, cultural, and experiential, of specific learners.

In response to these challenges, LDT researchers and practitioners traditionally have either revised extant instruments to capture a broader range of outcome measures or proposed amended or novel usability methods to make them more applicable within a learning context (Abulfaraj & Steele, 2020; Gladman et al., 2020; Lim & Lee, 2007; Oztekin et al., 2013; Yang et al., 2012). However, most of the developed methods, including instruments, suffer from methodological limitations (Lu et al., 2022). Given the increasing recognition of the importance of usability in educational and technology products, the need for reliable usability instruments such as survey scales and questionnaires is both urgent and timely.

## 3. Methods

The development of the Comprehensive Assessment of Usability Scale for Learning Technologies (CAUSLT) underwent an iterative process. Activities conducted during the process include, domain specification, item generation, validity checks, item refinement, and reliability checks. To demonstrate the version history of CAUSLT, a visual aid is provided in Figure 2. Specifically, the initial item generation resulted in Alpha 1.1, which underwent two evaluation activities for validity checks, namely, (1) the initial expert evaluation and (2) the think-aloud sessions. An item quality review was conducted following the validity checks. The findings from these activities guided the revision of Alpha 1.1, leading to the creation of Alpha 1.2, which was distributed to experts for a second round of review. Based on experts' feedback, minor changes were incorporated in Alpha 1.2, resulting in the next version of the instrument (Alpha 1.3). Alpha 1.3 was piloted on a representative sample for internal consistency and internal structure using a range of statistical analyses. A few more changes were made to Alpha 1.3 based on results of the statistical analyses, leading to the final version of CASULT (Beta 1.1), which was piloted again with a distinct sample.

**Figure 2**

*Version history of the Comprehensive Assessment of Usability Scale for Learning Technologies.*

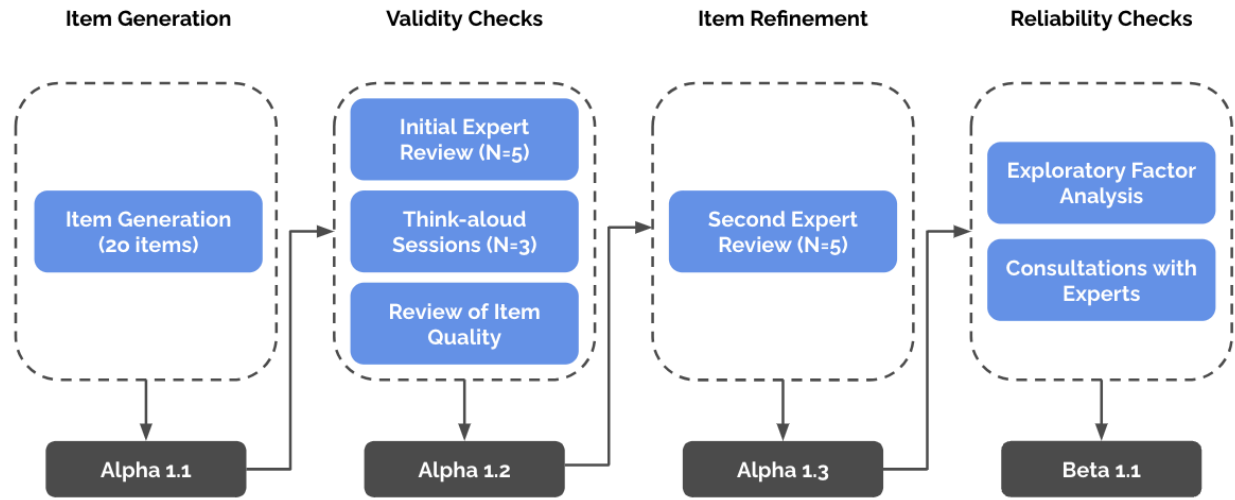

### 3.1 Usability Domain Specification

Usability in learning technologies is multifaceted, encompassing technological, pedagogical, and sociocultural aspects (Lu et al., 2022). Each aspect offers a unique lens through which the design and interaction of learning technologies can be evaluated, ensuring they cater to the diverse needs and contexts of learners.

#### 3.1.1 Technological usability

Technological usability refers to the ease with which users can interact with a UI to achieve specific goals related to the tool (cf. ISO, 2011, 2019; Nielsen, 1990), considering factors such as ease of use, efficiency, error frequency and severity, and user satisfaction. In the context of learning technologies, technological usability assesses how users (e.g., instructors, learners) navigate and interact with features of digital learning platforms. A system with high technological usability enhances the learner experience (LX) by making it intuitive and free from unnecessary barriers.

#### 3.1.2 Pedagogical usability

Pedagogical usability emphasizes the design and interaction within learning technologies to ensure they support learners in achieving their educational objectives (cf. Nokelainen, 2005).

It encompasses the tools, content, interface, and tasks, aligning them with pedagogical goals tailored to diverse learners across various contexts. While it shares certain elements with technical usability, such as user control and feedback, its primary focus is on balancing LX with the enhancement of long-term learning. This approach ensures that the learning technology and its embedded content effectively facilitate the attainment of learning goals set by both educators and learners.

### 3.1.3 Sociocultural usability

Sociocultural usability encompasses both the shared and the individual sociocultural contexts that influence sensemaking (Schmidt et al., 2024). Instead of focusing solely on aspects of social presence and group activities, sociocultural usability considers the lived experiences and cultural backgrounds of users (Smith & Ragan, 2004). It emphasizes the importance of understanding and designing for the sociocultural factors that shape how users interact with and perceive learning technologies. This perspective recognizes that learners come from diverse backgrounds, and their cultural and social experiences can significantly impact their interactions with learning technologies (Gray, 2020).

### 3.2 Item Generation and Reliability and Validity Analysis

A thorough review of literature was conducted, with a focus on usability instruments that have been applied to learning technologies. Following this, a content analysis was performed on the full corpus of collected instruments, specifically at an item level. Each item was examined to determine the outcome measure of usability. These outcome measures were then classified in accordance with the sociocultural, technical, and pedagogical framework. For instance, the outcome measure of the item "The color scheme of buttons is attractive" would be aesthetics, which represents a manifestation of the "Technological" subscale of usability. An initial set of 20 items was drafted and operationalized on a 7-point Likert type scale, following item writing principles identified from the literature (c.f. Clark & Watson, 2016; Nemoto & Beglar, 2014; Webb et al., 2002; Wolfe & Smith Jr., 2007).

The traditional instrument development process often involves starting with a large pool of initial items that are subsequently eliminated through various stages of refinement. While this approach can be effective in some contexts, we purposefully skipped this step because (1) an exhaustive review of the literature and existing instruments has been conducted, and (2) some dimensions (i.e., technological and pedagogical usability) have been well defined in the literature over the past 20 years, especially for the constructs that need to be assessed. This was because an exhaustive review of the literature was conducted, focusing on both the theoretical foundations of the concept of usability and the instruments that have been applied in the field of LDT. The response options range from strongly disagree to strongly agree, with a neutral point being neither agree nor disagree. The initial draft of CAUSLT was named Alpha 1.1, presented in Table 1.

**Table 1**

*Items constructed in CAUSLT Alpha 1.1, with associated outcome measures of usability, corresponding subscales, and sources.*

| Subscale | Item Content | Outcome Measures | Sources |
|---|---|---|---|
| Technological | The application was easy to use. | Ease of use | Horila et al., 2002; Nahm et al., 2006 |
| | I was able to finish tasks quickly without having to click or scroll excessively. | Efficiency | ISO 9241-210, 2019; ISO/IEC 25010, 2011 |
| | The application included features that made it easier to use for individuals with disabilities. | Accessibility | ISO 9241-210, 2019; ISO/IEC 25010, 2011 |
| | If I came across technical errors, I was able to fix them. | Error prevention & recovery | Nielsen, 1994a, 1994b; Squires & Preece, 1999 |
| | I like how the learning | Aesthetics | ISO 9241-210, 2019; |
| | application looks in general. Moving around to different sections of this application was straightforward. | Operability | Nielsen, 1994a, 1994b ISO 9241-210, 2019; ISO/IEC 25010, 2011 |
| | It was easy to learn how to use this application. | Learnability | Horila et al., 2002; ISO 9241-210, 2019 |
| | I was able to successfully finish tasks and scenarios with this application. | Usability | Horila et al., 2002; Nahm et al., 2006 |
| | I feel like I had control over what I wanted to do when using this application. | User control | ISO 9241-210, 2019; ISO/IEC 25010, 2011 |
| | Overall, I am satisfied with using the application. | Satisfaction | ISO 9241-210, 2019; ISO/IEC 25010, 2011 |
| Pedagogical | The learning materials provided by this application were of excellent quality. | Content | Lim & Lee, 2007; Moore et al., 2014; |
| | I felt fully engaged and absorbed while learning with this application. | Engagement | N/A |
| | I think this application was important and useful for reaching the learning goals. | Usefulness | Nokelainen, 2006 |
| | The application gave me feedback on how well I understood the learning materials. | Assessment | Jahnke, 2015; Moore et al., 2014 |
| | The learning materials in this application did not feel overwhelming to me. | Pedagogical strategy | Lim & Lee, 2007; Moore et al., 2014 |
| | The application provided enough support to help me reach the learning goals. | Support | Jahnke, 2015; Quinn, 1996; Silius et al., 2003 |
| | I could manage the speed and order of my own learning with this application. | Autonomy | Squires & Preece, 1999 |
| Sociocultural | I did not experience any prejudices based on culture, race, or gender when using the application/material. | Sociocultural sensitivity | N/A |

| | | |
|---|---|---|
| In my opinion, this application could be used by learners from diverse backgrounds. | Sociocultural sensitivity | N/A |
| Interacting and collaborating with others through this application aided in my own learning. | Sociocultural sensitivity | Nokelainen, 2006; Khanum & Trivedi, 2013 |

### 3.1.1 Usability Item Evaluation

The initial set of items was evaluated by five usability experts in the field of LDT for content validity using the Content Validity Index (Polit & Beck, 2006). The items were segmented as individual entries with a scale of 1 to 4 to avoid neutral or ambivalent midpoint (Lynn, 1986)). Specifically, a rating of 1 indicated that the item was "not relevant/not clear", 2 indicated that it "need some revision", 3 indicated that the item was "relevant/clear but need minor revision", and 4 was "very relevant/clear". A total of two rounds of expert evaluation were conducted.

To assess the response process validity, three undergraduate students taking educational technology or related classes at a major university located in the southeast region were recruited. Two of them were female native speakers of English, and the male participant's first language was Chinese (he was fluent in English for reading, writing, and communication in general). Individual think-aloud sessions were conducted using the online conferencing software Zoom. An informed consent was obtained from each participant prior to each session. During each session, items were displayed on the screen and shared with the participant. The participant was instructed to carefully review each item and respond to the question "What do you think this item asks?" Specifically, the participant was required to think-aloud, allowing the researcher to understand and take notes of the participant's natural thought process. Immediately after reviewing each item, the participant was asked if they had any additional comments or feedback regarding the item. All responses were documented in a spreadsheet.

Additionally, an evaluation of the general item quality was conducted using the matrix developed by Crocker and Algina (1986). The matrix provides a framework for assessing the item writing quality, taking into consideration factors such as clarity, conciseness, relevance, and appropriateness of the language used. This evaluation provided an additional layer of quality assurance for the developed items beyond the expert evaluation and think-aloud sessions.

### 3.1.2 Analysis Procedures

Two types of CVI scores (Yusoff, 2019) were calculated to assess the content validity of the instrument, item-level CVI (I-CVI) and scale-level CVI (S-CVI). The relevance ratings provided by experts were first coded into binary values which allowed for a simplified calculation of the CVI scores. A rating of 1 or 2 was coded as 0, and a rating of 3 or 4 was coded as 1. The I-CVI score was computed for each item by summing up the coded relevance ratings for that item from all experts involved in the rating process. The sum was then divided by the total number of experts. Items with an I-CVI score lower than the threshold of 0.78 were revisited for revision or removal, as suggested by Shi and colleagues (2012). The S-CVI score was computed with reference to the average (S-CVI/Ave) method (Yusoff, 2019). A score above 0.9 for S-CVI/Ave indicates excellent content validity (Shi et al., 2012).

### 3.2 Item Refinement and Reliability Analysis

Based on the feedback from the two rounds of expert evaluation, some items were reworded, some were entirely removed, and a few items were added to the instrument. The process to compute the I-CVI and S-CVI scores for both rounds of expert evaluation were identical.

### 3.2.1 Data Collection

The internal consistency of CAUSLT was checked using coefficient alpha, also known as Cronbach's Alpha (Cronbach, 1951).

***Participants and procedures.*** For this study, two rounds of data collection were conducted because EFA was conducted twice to fully examine the factor structure of the developed instrument. The initial round recruited 155 participants based on specific criteria through the Prolific platform, an online platform renowned for its diverse pool of registered participants who willingly sign up for various research studies. Participants were required to be located exclusively in the United States of America, aged between 18 and 30, and have English as their first language. Additionally, participants must have had a history of at least 20 approved studies in their prior experiences on the platform. Of the 155 participants, 46 were female, 108 were male, and one preferred not to say. In terms of race, 27 were Asian, 14 were Black, 23 were Mixed, nine were Others, and 82 were White. The second data collection recruited a total of 248 participants from multiple higher education institutions.

A desktop-based virtual reality (VR) training session (Figure 3) developed using learning experience design methods and processes (Schmidt et al., 2024) was utilized as the learning technology in this study. The training session consists of three micro-learning sessions and a navigation quest which serves as a formative assessment. It was designed for users who wish to learn to navigate desktop-based VR space. Upon successful completion, learners will be able to use a keyboard and a mouse to complete basic navigation (e.g., move forward/backward, turn, look up/down).

**Figure 3**

*A screenshot of the overview of the training session.*

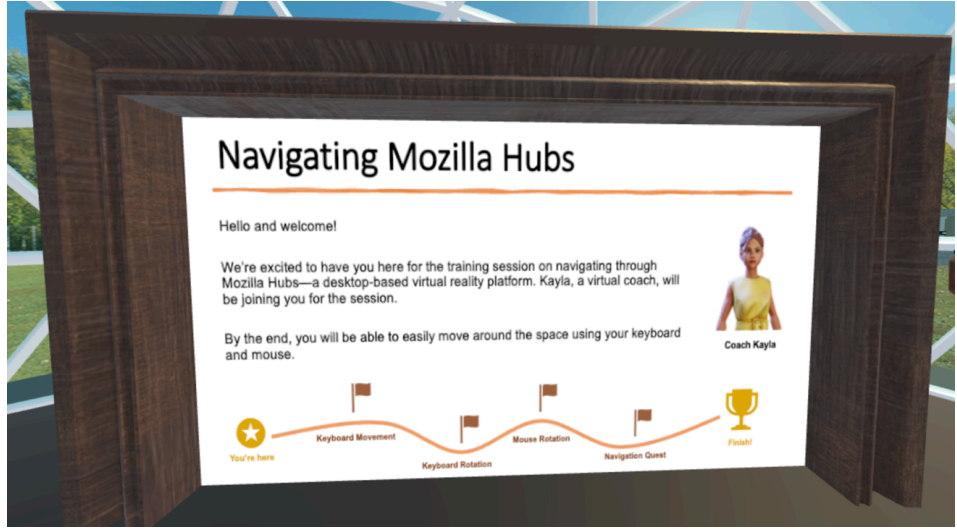

All data were collected online using Qualtrics. The link to the Qualtrics survey was embedded in a 3D trophy in the last virtual room in the training session (Figure 4). When participants hover over the trophy, an option of "open link" appears and the Qualtrics survey opens in a separate browser window upon on-click. Text-based instructions on how to access the survey was also provided in the training session.

**Figure 4**

*A screenshot of the last room where the link to the survey was embedded in the trophy.*

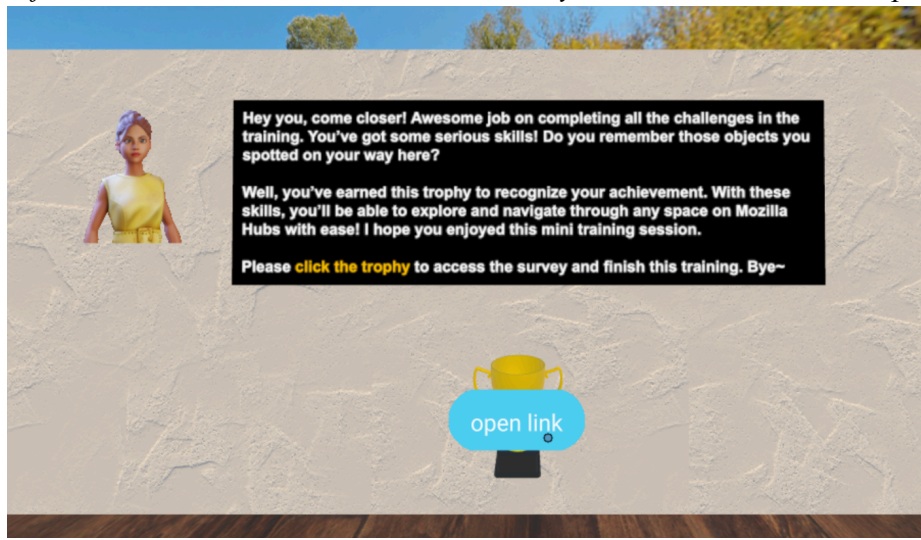

**3.2.2 Exploratory Factor Analysis**

The current study conducted two rounds of Exploratory Factor Analysis (EFA) to identify and examine the underlying factors that explain the correlations among the variables (Kim, 2008). Participants' responses to the instrument were imported into a spreadsheet for further analysis. Ordinal, Likert-scale ratings (e.g., strongly agree) were transformed into numerical data using the built-in feature in Qualtrics. All data were reverse coded and all analyses were performed in R Studio (v4.3.0; R Core Team, 2023), with the specific packages including, psych (v2.3.3; Revelle, 2023), ltm (v1.2.0; Rizopoulos, 2006), semTools (v0.5.6; Jorgensen et al, 2022), and lavaan (v0.6.15; Rosseel, 2012).

Procedures to run both EFA analyses were identical but with distinct samples. Assumptions of sample adequacy, sphericity, normality, linearity, and factorability were checked prior to each analysis. Three methods were used in conjunction to make an informed decision on the number of factors to retain: (1) Kaiser's criterion, also known as the K1 method (Kaiser, 1958), (2) a scree plot, and (3) the parallel analysis (Horn, 1965). The Principal Axis Factoring (PAF) estimation method followed by the Promax rotation was used in both analyses. The communality cutoff threshold was set to .4 ($\lambda = .4$) according to the recommendation of Costello and Osborne (2019) when using the PAF estimator along with the Promax rotation.

## 4. Results

The development of CAUSLT underwent an iterative process. Results are reported for the following activities: (1) initial expert evaluation, (2) think-aloud sessions, (3) changes incorporated in the instrument, (4) second expert evaluation, and (5) two rounds of exploratory factor analysis.

**4.1 The Initial Expert Evaluations**

As suggested by Shi and colleagues (2012), items with an I-CVI score lower than the threshold of 0.78 should be revisited for revision or removal. The I-CVI scores from the first expert evaluation indicated that 11 out of the 20 items were in need of revision. The S-CVI/Ave score was .67. Given that excellent content validity is characterized in the literature as having S-CVI/Ave of .9 or higher (Shi et al., 2012), the Alpha 1.1 version did not meet the established threshold for acceptability.

Two experts provided additional comments regarding the overall quality of Alpha 1.1. One expert expressed concerns on items that assessed the pedagogical and sociocultural subscales of usability, as they were written in a way such that they were largely dependent on educators implementing the learning technology and the evaluator/end users themselves. Another expert made a suggestion to add one or two questions on collaboration.

**4.2 Think-aloud Sessions**

In an ideal situation, participants' responses to each question should align with the item's intended construct of usability. This was true for most items, as participants were able to identify the underlying construct, although using their own words and not necessarily using the exact construct terminology. For example, the item "The application was easy to use" measured the ease-of-use. All participants recognized the usability outcome measure, indicating the clearness and quality of it. This finding was consistent with the results in the initial round of expert evaluation, where the I-CVI score for this item was 1, indicating that all experts agreed it was clear and relevant.

**4.3 Changes Incorporated in Alpha 1.1**

Based on the results from the initial round of expert evaluation and the think-aloud sessions, all items were revisited for clarity, relevance, consistency, and generalizability, leading to the creation of Alpha 1.2. A comparison of the original items and updated items was provided in Appendix A. Additionally, a column was added to indicate the type of changes made, i.e., no change, revised, removed, and newly added.

### 4.4 The Second Expert Evaluation

The I-CVI scores improved for all items (with the exception of those with original I-CVI scores of 1). The S-CVI/Ave score was .9, indicating excellent content validity. With the feedback from the experts, only minor revisions were made to two items in Alpha 1.2 with reference to comments from experts, resulting in the next version, Alpha 1.3 (Appendix B), which was used for the initial EFA.

### 4.5 Two Rounds of Exploratory Factor Analysis

Due to the exploratory nature of the current study, we conducted EFA twice with distinct samples to obtain the most appropriate factor structure.

#### 4.5.1 Factor structure from the initial EFA

All assumptions for conducting an EFA held except for normality, which was not uncommon for ordinal data. Results from the three extraction methods were consistent in that a small number of factors (the first three factors) could explain a large portion of the variance in the data. Specifically, a three-factor structure was selected. The coefficient alpha of the Alpha 1.3 version of the instrument was .934, indicating a high level of internal consistency (Nunnally, 1978) among the 21 variables in the dataset. The coefficient alpha for each factor and loadings for variables are recorded in Appendix C. Within this model, variables 3 ("In my opinion, this application is accessible for individuals with disabilities") and 4 ("If I came across technical errors, I was able to fix them") did not exhibit a satisfactory loading on any of the factors. A number of issues could have contributed to the weak loading, including item redundancy and conceptual mismatch. Furthermore, variable 18 ("I can manage the pace of my learning with this application") cross-loaded on Factor 2 and Factor 3 with similar loadings. Results indicated the need for potential revision or removal of variables 3, 4 and 18.

In summary, the initial factor structure was not in perfect alignment with the conceptual framework. Results of the initial analysis guided the revision of Alpha 1.3, leading to the final version, Beta 1.1 of CAUSLT (Table 2).

### 4.5.2 Second round of EFA

Our team executed a series of pivotal actions. We first assessed factor loadings for each variable in the three-factor model. Next, we engaged the expertise of a psychometrician for valuable insights. Following this, our research's theoretical framework was revisited for relevance. Finally, we consulted with an expert in usability research, learning experience design, and HCI. In addition to it, a minor design change was implemented in the selected learning technology.

**Table 2**

*Item content in the final version of CAUSLT.*

| Item # | Items in Beta 1.1 |
|---|---|
| 1 | This application is easy to use. |
| 2 | I am able to complete tasks quickly when using this application. |
| 3 | In my opinion, this application could be used by anyone, regardless of their physical, cognitive, and sensory abilities. |
| 4 | This application is helpful in recovering from errors, if encountered. |
| 5 | Navigating within this application is straightforward. |
| 6 | It is easy to learn how to use this application. |
| 7 | I feel like I have control over what I need to do when using this application. |
| 8 | I like the look-and-feel of this application. |
| 9 | This application has all the functions I expect it to have for achieving my learning goals. |
| 10 | Overall, I am satisfied with this application. |
| 11 | Content provided by this application is of excellent quality. |
| 12 | Content in this application does not feel overwhelming to me. |
| 13 | I feel engaged while learning with this application. |
| 14 | I think this application is useful for reaching my learning goals. |
| 15 | This application provides feedback on how well I understand the content. |
| 16 | This application provides enough support in my learning process. |
| 17 | I can manage the pace of my learning with this application. |
| 18 | I do not experience prejudices based on culture, race, gender, or other factors when using this application. |
| 19 | This application provides opportunities that promote social interaction among learners. |
| 20 | In my opinion, this application could be used by a range of learners, regardless of their social, cultural, or demographic backgrounds. |

The final version of CAUSLT consists of 20 items that assess three subscales of usability. Specifically, the first 10 items assessed the pedagogical subscale of usability, including ease-of-use, learnability, efficiency, and more. The following seven items assessed the technological subscale of usability, including information quality, engagement, usefulness, and more. The last three items assessed the sociocultural subscale of usability, including social interaction and diversity, equity, and inclusion.

All assumptions of conducting EFA were met except for the normality assumption. The Kaiser-Meyer-Olkin (KMO) measure was excellent at 0.95, indicating that a significant proportion of variance might be common among the variables. Bartlett's Test of Sphericity returned a chi-square value of 5147.295 with a p-value of approximately 0, confirming that the correlation matrix significantly differs from the identity matrix and therefore supporting the factorability of the dataset. Cronbach's Alpha for the dataset is extremely high at 0.95, suggesting excellent internal consistency among the items. The coefficient alpha for each factor and loadings for variables are recorded in Table 3. A factor diagram is provided in Appendix D.

**Table 3**

*Loadings of the rotated factors with items in the final version of CAUSLT and their alpha coefficient.*

| | Factor 1 | Factor 2 | Factor 3 |
|---|---|---|---|
| **Factor 1 – Pedagogical** | $\alpha = 0.946$ | | |
| V14. I think this application is useful for reaching my learning goals. | 1.016[1] | | |
| V8. I like the look-and-feel of this application. | 0.846 | | |
| V11. Content provided by this application is of excellent quality. | 0.845 | | |
| V9. This application has all the functions I expect it to have for achieving my learning goals. | 0.798 | | |
| V15. This application provides feedback on how well I understand the content. | 0.788 | | |
| V16. This application provides enough support in my learning process. | 0.765 | | |
| V10. Overall, I am satisfied with this application. | 0.648 | | |
| V13. I feel engaged while learning with this application. | 0.594 | | |
| V4. This application is helpful in recovering from errors, if encountered. | 0.585 | | |
| V17. I can manage the pace of my learning with this application. | 0.416 | | |
| **Factor 2 – Technological Usability** | $\alpha = 0.923$ | | |
| V2. I am able to complete tasks quickly when using this application. | | 0.91 | |
| V6. It is easy to learn how to use this application. | | 0.884 | |
| V1. The application is easy to use. | | 0.884 | |
| V5. Navigating within this application is straightforward. | | 0.812 | |
| V7. I feel like I have control over what I need to do when using this application. | | 0.663 | |
| V12. Content in this application does not feel overwhelming to me. | | 0.62 | |
| V3. In my opinion, this application could be used by anyone, regardless of their physical, cognitive, and sensory abilities. | | 0.501 | |
| **Factor 3 – Sociocultural Usability** | $\alpha = 0.82$ | | |
| V18. I do not experience prejudices based on culture, race, gender, or other factors when using this application. | | | 0.847 |
| V20. In my opinion, this application could be used by a range of learners, regardless of their social, cultural, or demographic backgrounds. | | | 0.804 |
| V19. This application provides opportunities that promote social interaction among learners. | | | 0.738 |

[1]*The loading of this variable on Factor 1 was greater than 1 because a polychoric correlation matrix was created, which was more appropriate for analyzing ordinal data than using Pearson's correlations. A pattern loading can lie beyond [-1, 1].*

Our team is preparing data to perform a confirmatory factor analysis (CFA) using Beta 1.1 to confirm the factor structure. Model fit indices will be computed for the three-factor model and its variations. Due to the exploratory nature of the current study, we will report detailed results of the CFA in the future.

## 5. Discussion and Implications

A thorough content analysis was performed that informed the development of CAUSLT, followed by a range of statistical analyses to assess its psychometric properties. The overarching research question that guided these analyses was, what is the underlying factor structure of the developed usability survey instrument? In this section, we discuss the analysis of CAUSLT, followed by its alignment to theories in the field.

### 5.1 Analysis of CAUSLT

CAUSLT was designed and developed with reference to the literature, including the STP framework (Lu et al., 2022; Jahnke et al., 2020), the role of culture in usability evaluation (Downey et al., 2005; Ford & Kotzé, 2005), and existing usability instruments that have been applied in the field of LDT. Following a rigorous procedure for instrument development (Bandalos, 2018; Clark & Watson, 2016), the final version of CAUSLT achieved excellent overall reliability ($\alpha = 0.95$), with satisfactory to excellent reliability across its subscales (see Table 3). Items comprising each factor demonstrate a strong alignment with the STP framework, except for items 4 and 12, which are discussed in the following.

Items loaded on Factor 1 (Pedagogical Usability) primarily focus on how the learning technology supports the learners' perceived learning goals and engagement with the content, except Item 4 ("This application is helpful in recovering from errors, if encountered"). This item was intended to measure "error recovery", a trait of the technological usability (Lu et al., 2022) but ended up loading sufficiently on the pedagogical factor ($\lambda = .059$). Several reasons could explain the off loading of this item, including the lack of relevance and variability in user expertise. The lack of relevance refers to the scenario in which the item was not perceived as highly relevant by the respondents in the study, leading to limited attention to this item. In this study, respondents' expertise in VR technology, including familiarity, could have influenced their responses to this item. If respondents were highly skilled and experienced in desktop-based VR technologies, they might have encountered technical errors less frequently, or they might be able to resolve technical issues without relying on external help, therefore not considering this item as relevant. On the contrary, novice respondents in this type of learning technology might have struggled to fix technical errors, therefore finding the item difficult to respond to. As we anticipate more pilot testing using different learning technologies, such as online learning modules, we will continue to examine the performance of this item.

The Technological Usability factor consists of items that ensure learners do not feel hindered by the interface or functionality of the learning technology. The significant loading of Item 12 ("Content in this application does not feel overwhelming to me") on this factor ($\lambda = .62$) rather than the pedagogical factor was unexpected, but overall not surprising, as technological simplicity can reduce cognitive overload (Feldman, 2016). Additionally, it speaks to the nuanced interactions between content presentation and how it may affect learners' perception with the given learning technology. As shown in the factor structure (see Appendix D), factors show moderate to high correlations with each other, indicating that while they represent distinct dimensions, these dimensions are related. This finding is consistent with the STP framework (Jahnke et al., 2020), which guided the development of CAUSLT. Alternatively, this grouping could be caused by respondents' misinterpretation of what the word "content" implies. From the perspective of item construction, by content, we referred to the learning materials designed using sound instructional strategies, while the learners in the context of the current study could have simply perceived it as the layout of information in the VR environment. More data would be needed to determine the root cause of the loading of this variable.

Factor 3 (Sociocultural Usability) aims to assess the inclusiveness and social impact of the learning technology. Although the reliability within this factor was relatively low ($\alpha = 0.82$) compared to the first two factors, it was considered acceptable (Nunnally, 1978).

In summary, the analysis of CAUSLT revealed a robust underlying factor structure, with three meaningful factors each representing a distinct but still related underlying dimension of the data.

**5.2 Theoretical Alignment**

CAUSLT was developed to address a critical gap in usability research in the field of LDT that learning technologies should be evaluated from multiple perspectives. Studies have consistently shown that traditional usability metrics and instruments (e.g., ISO standards and CSUQ), while sufficient for evaluating general information and communication technologies, do not fully capture the unique features of learning technologies. The final round of analysis revealed a three-factor structure which successfully replicated our conceptual framework. This supports research from the past decade that suggests learning is enhanced by pedagogical strategies and embedded in social contexts. Therefore, in order to effectively evaluate a learning technology that is designed to promote learning, it is important to consider its multifaceted usability instead of focusing solely on one dimension.

An example that connects technological usability (e.g., accessibility) and pedagogical usability (e.g., instructional strategy) is the prominence of inclusive design in the fields of HCI and LDT, aiming to address the needs of vulnerable and underrepresented populations using different approaches to accommodate local, cultural, political, and historical situations (Keates et al., 2000). From a technological perspective, ease-of-use and accessibility are fundamental attributes emphasized in traditional usability standards such as ISO 9241 and ISO/IEC 25010. These standards highlight the importance of considering users' needs and capabilities when designing and developing interactive systems (Dorrington et al., 2016). From a pedagogical perspective, instructional strategies such as Universal Design for Learning (UDL; Rose, 2001) and Differentiated Instruction (Tomlinson et al., 2003), prioritize ease-of-use and accessibility. Specifically, UDL principles guide the overall design of instruction to ensure accessibility and flexibility, while Differentiated Instruction allows for personalized instruction tailored to individual needs, capabilities, and diverse backgrounds. Taken together, the accessibility of learning technologies should not be assumed as inherent qualities, rather, they are the outcomes of purposeful design considerations and features grounded in instructional strategies (Moore & Ellsworth, 2014). By acknowledging the diversity of users and seeking solutions that make designs functional for diverse user groups, researchers can ensure that learning technologies are accessible and adaptable to the needs of as many users as possible (Moore & Ellsworth, 2014).

  Cultural usability is another important consideration that recognizes the impact of culture on UX and usability (Clemmensen, 2009; Vatrapu, 2008). Usability testing outcomes can differ across Western and Eastern cultural perspectives (Yeo, 1998), underscoring the need to provide content and features that are relevant and accessible to users from diverse cultural backgrounds. Moreover, instructional strategy and sociocultural sensitivity contribute to the ease-of-use and accessibility of learning technologies, by considering learners' cognitive abilities, levels of technology literacy, and diverse backgrounds, ultimately enhancing the overall usability of the designed learning technology. Oftentimes viewed together with cultural usability is the social aspect of usability. Illustrated in the work by Whitbeck (1996), the tangible effectiveness of accessible design extends to favorable societal outcomes in the realm of education, leading to the reduction of obstacles and enhanced access for historically underserved or marginalized groups of learners.

Taken together, these findings not only explain the correlated relationship among the three factors, but also reinforce the interconnected nature of the three dimensions of usability.

**5.3 Practical Implication**

The development of CAUSLT is a direct response to the new conceptualization of usability for learning technologies (Lu et al., 2022, Schmidt et al., 2024). It expands from representing and assessing only one aspect of usability (i.e., the technological aspect assessed by SUS) to including learning-related aspects (i.e., pedagogical and sociocultural aspects). In practice, this instrument can significantly impact the design, development, and implementation of learning technologies by offering detailed insights into various usability aspects for researchers and practitioners. When adopted early and iteratively in a developmental cycle, usability evaluation enables designers to identify and address usability issues across multiple dimensions promptly (Lyzara et al., 2019), potentially enhancing the overall learning experience. As technology-mediated learning continues to evolve in our field, the quality of learning experience directly influences learning outcomes.

Methodologically, the development of CAUSLT serves as an example of following sound and rigorous procedures in instrument development to establish adequate validity and reliability, which are critical properties of an effective instrument, yet oftentimes overlooked in self-developed usability tools (Quiñones et al., 2018). The results of a usability evaluation are only valid if there are no methodological flaws, such as the use of unvetted instruments.

## 5.4 Limitations

A few limitations should be considered in conjunction with the findings. Firstly, practical limitations limited the recruitment of participants for the think-aloud sessions aiming to assess response-process validity. As a result, only three students with background knowledge in educational technology were included in the study. Out of the three participants, only one participant was a non-native speaker. This small sample may have implications for the quality and generalizability of the feedback obtained, particularly in relation to items assessing the sociocultural subscale of usability.

Secondly, the sampling method used in this study could result in a non-representative sample, as participants on Prolific are typically self-selected and incentive-motivated to participate in research studies. Similarly, the demographic characteristics and background of participants may slightly differ from the target populations of interest (i.e., users of learning and educational technologies). An additional concern from sampling on Prolific relates to response bias, because participants may have specific response patterns, ranging from highly interested in research studies in general to passively engaged due to the compensation. All of this could limit the generalizability of the findings, especially when making population-based inferences.

## 5.5 Future Research

When reviewing the results from the statistical analyses alongside the existing literature, participants' prior experience or exposure to interactive technologies tended to impact their perceived usability (Hurtienne et al., 2013). To further investigate and better discern the influence of learners' prior experience on the perceived usability, it is recommended that researchers and practitioners administer a short technology competence survey prior to the administration of CAUSLT. This combined approach would enable a more comprehensive exploration of the relationship between users' prior experience and the perceived usability.

As mentioned earlier, our research team is actively piloting CAUSLT Beta 1.1 with different and diverse populations to evaluate measurement invariance. This will be done by using the Multiple Indicator Multiple Cause (MIMIC) model, which examines whether the observed variables adequately represent the latent construct and whether they are influenced by external factors. Specifically, it helps understand if the three-factor structure between the items and the latent construct is consistent across the subgroups being compared.

Last but not least, our research team will be focusing on establishing benchmarks for CAUSLT. Relevant criteria based on existing literature, expert opinions, and specific thresholds that indicate satisfactory performance (e.g., metrics for system effectiveness and efficiency) might be used for this purpose. We believe CAUSLT provides a useful tool for advancing usability research in the field of LDT. We encourage researchers to adopt the CAUSLT instrument and welcome future collaborations, as its use and continuous refinement through critical discourse could lead to superior LX with learning technologies, promoting higher standards of usability and learning effectiveness.

## 6. Conclusion

This paper detailed an iterative and rigorous process of the design, development, and evaluation analysis of CAUSLT–the first survey instrument in the field of LDT that considers three aspects of usability of a given learning technology. Results of the final round of EFA were discussed in detail with reference to the literature, highlighting the findings of the three-factor model and its implications. It is important to emphasize the significant impact this instrument can have on usability research in our field. The reliability of the CAUSLT makes it a valuable tool for assessing the multifaceted usability of learning technologies. Its incorporation into future research endeavors can greatly enhance the understanding of usability of the evaluated technologies and address corresponding challenges to promote learning. By providing a reliable and standardized measure, CAUSLT empowers researchers and practitioners to make informed decisions and improvements in the design and implementation of learning technologies, ultimately leading to enhanced learner experiences.

# References

Abulfaraj, A., & Steele, A. (2020). Coherent Heuristic Evaluation (CoHE): Toward Increasing the Effectiveness of Heuristic Evaluation for Novice Evaluators. In A. Marcus & E. Rosenzweig (Eds.), *Design, User Experience, and Usability. Interaction Design* (Vol. 12200, pp. 3–20). Springer International Publishing. https://doi.org/10.1007/978-3-030-49713-2_1

Amant, K. S. (2017). *The Cultural Context of Care in International Communication Design: A Heuristic for Addressing Usability in International Health and Medical Communication*. 9.

Bartlett, M. S. (1950). Tests of significance in factor analysis. British Journal of Psychology, 3, 77–85.

Bender, T., Huesmann, R., & Heinemann, A. (2021). Software Development Processes for ADs, SMCs and OSCs supporting Usability, Security, and Privacy Goals – an Overview. *The 16th International Conference on Availability, Reliability and Security*, 1–6. https://doi.org/10.1145/3465481.3470022

Brooke, J. (1996). System usability scale (SUS): A quick-and-dirty method of system evaluation user information. *Reading, UK: Digital Equipment Co Ltd*, *43*, 1–7.

Cayola, L., & Macías, J. A. (2018). Systematic guidance on usability methods in user-centered software development. *Information and Software Technology*, *97*, 163–175. https://doi.org/10.1016/j.infsof.2018.01.010

Clark, L. A., & Watson, D. (2016). Constructing validity: Basic issues in objective scale development.

Clemmensen, T. (2009). Towards a Theory of Cultural Usability: A Comparison of ADA and

CM-U Theory. In M. Kurosu (Ed.), *Human Centered Design* (Vol. 5619, pp. 416–425). Springer Berlin Heidelberg. https://doi.org/10.1007/978-3-642-02806-9_48

Costabile, M. F., De Marsico, M., Lanzilotti, R., Plantamura, V. L., & Roselli, T. (2005). On the usability evaluation of e-learning applications. In *Proceedings of the Annual Hawaii International Conference on System Sciences* (p. 6). https://www.scopus.com/inward/record.uri?eid=2-s2.0-27544464911&partnerID=40&md5=aa8e0a7354ed9217d69cf5b7f5670767

Costello, A. B., & Osborne, J. (2019). Best practices in exploratory factor analysis: Four recommendations for getting the most from your analysis. *Practical assessment, research, and evaluation*, *10*(1), 7.

Crocker, L., & Algina, J. (1986). *Introduction to Classical and Modern Test Theory*. Holt, Rinehart and Winston, 6277 Sea Harbor Drive, Orlando, FL 32887 ($44.75). https://eric.ed.gov/?id=ed312281

Cronbach, L. J. (1951). Coefficient alpha and the internal structure of tests. *Psychometrika*, *16*(3), 297–334. https://doi.org/10.1007/BF02310555

Curcio, K., Santana, R., Reinehr, S., & Malucelli, A. (2019). Usability in agile software development: A tertiary study. *Computer Standards & Interfaces*, *64*, 61–77. https://doi.org/10.1016/j.csi.2018.12.003

Devy, N. P. I. R., Wibirama, S., & Santosa, P. I. (2017). Evaluating user experience of english learning interface using User Experience Questionnaire and System Usability Scale. *2017 1st International Conference on Informatics and Computational Sciences (ICICoS)*, 101–106. https://doi.org/10.1109/ICICOS.2017.8276345

Diaz, V., Albritton, T., Katague, M., Dancy, V., Breny, J., & Kershaw, T. (2013). We Are Family:

Effects of a Relationship-Strengthening Prevention Intervention on Parenting Behaviors Among Black and Latino Adolescent Couples. *JOURNAL OF FAMILY ISSUES*. https://doi.org/10.1177/0192513X211064860

Dorrington, P., Wilkinson, C., Tasker, L., & Walters, A. (2016). User-centered design method for the design of assistive switch devices to improve user experience, accessibility, and independence. Journal of Usability Studies, 11(2).

Downey, S., Wentling, R. M., Wentling, T., & Wadsworth, A. (2005). The relationship between national culture and the usability of an e-learning system. *Association for the Advancement of Computing in Education*, *6*(1), 119–146. https://doi.org/10/fjs767

Feldman, J. (2016). The simplicity principle in perception and cognition. *Wiley Interdisciplinary Reviews: Cognitive Science*, *7*(5), 330-340.

Ford, G., & Kotzé, P. (2005). Designing Usable Interfaces with Cultural Dimensions. In M. F. Costabile & F. Paternò (Eds.), *Human-Computer Interaction—INTERACT 2005* (Vol. 3585, pp. 713–726). Springer Berlin Heidelberg. https://doi.org/10.1007/11555261_57

Ford, G., Kotzé, P., & Marcus, A. (2005). Cultural dimensions: Who is stereotyping whom. *Proceedings of the 11th International Conference on Human-Computer Interaction*.

Gamage, D., Perera, I., & Fernando, S. (2020). MOOCs Lack Interactivity and Collaborativeness: Evaluating MOOC Platforms. *International Journal of Engineering Pedagogy (IJEP)*, *10*(2), 94. https://doi.org/10/ggwf3w

Gladman, T., Tylee, G., Gallagher, S., Mair, J., Rennie, S. C., & Grainger, R. (2020). A Tool for Rating the Value of Health Education Mobile Apps to Enhance Student Learning (MARuL): Development and Usability Study. *JMIR MHealth and UHealth*, *8*(7), e18015. https://doi.org/10.2196/18015

Granić, A., & Ćukušić, M. (2011). Usability Testing and Expert Inspections Complemented by Educational Evaluation: A Case Study of an e-Learning Platform. *Educational Technology & Society*, *14*(2), 107–123.

Gray, C. M. (2020). Paradigms of knowledge production in human-computer interaction: Towards a framing for learner experience (LX) design. M. Schmidt, AA Tawfik, I. Jahnke, & Y. Earnshaw,(2020). Learner and User Experience Research: An Introduction for the Field of Learning Design & Technology. EdTech Books. https://edtechbooks.org/ux.

Hall, E. T. (1973). *The Silent Language*. Knopf Doubleday Publishing Group.

Heldal, I. (2007). *The Impact of Social Interaction on Usability for Distributed Virtual Environments*. 11.

Hofstede, G. (2011). Dimensionalizing Cultures: The Hofstede Model in Context. *Online Readings in Psychology and Culture*, *2*(1). https://doi.org/10.9707/2307-0919.1014

Holthe, T., Halvorsrud, L., Karterud, D., Hoel, K.-A., & Lund, A. (2018). Usability and acceptability of technology for community-dwelling older adults with mild cognitive impairment and dementia: A systematic literature review. *Clinical Interventions in Aging*, *13*, 863–886. https://doi.org/10.2147/CIA.S154717

Horn, J. L. (1965). A rationale and test for the number of factors in factor analysis. *Psychometrika*, *30*(2), 179–185. https://doi.org/10.1007/BF02289447

Hurtienne, J., Horn, A.-M., Langdon, P. M., & Clarkson, P. J. (2013). Facets of prior experience and the effectiveness of inclusive design. *Universal Access in the Information Society*, *12*(3), 297–308. https://doi.org/10.1007/s10209-013-0296-1

ISO. (2011). ISO/IEC 25010:2011(en), Systems and software engineering—Systems and software Quality Requirements and Evaluation (SQuaRE)—System and software quality models. (n.d.). Retrieved September 9, 2021, from https://www.iso.org/obp/ui/#iso:std:iso-iec:25010:ed-1:v1:en:sec:4.4

ISO. (2019). ISO 9241–210:2019(en), Ergonomics of human-system interaction—Part 210: Human-centred design for interactive systems. Retrieved September 9, 2021, from https://www.iso.org/obp/ui/#iso:std:iso:9241:-210:ed-2:v1:en

Jahnke, I., Schmidt, M., Pham, M., & Singh, K. (2020). Sociotechnical-pedagogical usability for designing and evaluating learner experience in technology-enhanced environments. Learner and user experience research, 127-145.

Jorgensen, T. D., Pornprasertmanit, S., Schoemann, A. M., & Rosseel, Y. (2022). semTools: Useful tools for structural equation modeling. R package version 0.5-6. Retrieved from https://CRAN.R-project.org/package=semTools

Kaiser, H. F. (1958). The varimax criterion for analytic rotation in factor analysis. *Psychometrika*, *23*(3), 187–200. https://doi.org/10.1007/BF02289233

Keates, S., Clarkson, P. J., Harrison, L. A., & Robinson, P. (2000, November). Towards a practical inclusive design approach. In Proceedings on the 2000 conference on Universal Usability (pp. 45-52).

Kim, H.-J. (2008). Common Factor Analysis Versus Principal Component Analysis: Choice for Symptom Cluster Research. *Asian Nursing Research*, *2*(1), 17–24. https://doi.org/10.1016/S1976-1317(08)60025-0

Kirakowski, j. (1996). The Software Usability Measurement Inventory: Background and Usage. In *Usability Evaluation In Industry*. CRC Press.

Lachner, F., von Saucken, C., 'Floyd' Mueller, F., & Lindemann, U. (2015). Cross-Cultural User

Experience Design Helping Product Designers to Consider Cultural Differences. In P. L. P. Rau (Ed.), *Cross-Cultural Design Methods, Practice and Impact* (pp. 58–70). Springer International Publishing. https://doi.org/10.1007/978-3-319-20907-4_6

Lim, C. J., & Lee, S. (2007). Pedagogical Usability Checklist for ESL/EFL E-learning Websites. *Journal of Convergence Information Technology*, *2*(3), 10.

Lu, J., Schmidt, M., Lee, M., & Huang, R. (2022). Usability research in educational technology: A state-of-the-art systematic review. Educational technology research and development, 70(6), 1951-1992.

Lynn, M. R. (1986). Determination and Quantification Of Content Validity. *Nursing Research*, *35*(6), 382.

Lyzara, R., Purwandari, B., Zulfikar, M. F., Santoso, H. B., & Solichah, I. (2019). E-government usability evaluation: Insights from a systematic literature review. *In Proceedings of the 2nd International Conference on Software Engineering and Information Management* (pp. 249-253).

Maramba, I., Chatterjee, A., & Newman, C. (2019). Methods of usability testing in the development of eHealth applications: A scoping review. *International Journal of Medical Informatics*, *126*, 95–104. https://doi.org/10.1016/j.ijmedinf.2019.03.018

Mardia, K. V. (1980). 9 Tests of unvariate and multivariate normality. In Handbook of Statistics (Vol. 1, pp. 279–320). Elsevier. https://doi.org/10.1016/S0169-7161(80)01011-5

Marell-Olsson, E., & Jahnke, I. (2019). Wearable Technology in a Dentistry Study Program: Potential and Challenges of Smart Glasses for Learning at the Workplace. In *Perspectives on Wearable Enhanced Learning (WELL)* (pp. 433–451). Springer.

Moon, N. W., Baker, P. M., & Goughnour, K. (2019). Designing wearable technologies for users

with disabilities: Accessibility, usability, and connectivity factors. *Journal of Rehabilitation and Assistive Technologies Engineering*, 6, 2055668319862137. https://doi.org/10.1177/2055668319862137

Moore, J. L., Dickson-Deane, C., & Liu, M. Z. (2014). Designing CMS courses from a pedagogical usability perspective. In A. D. Benson (Ed.), *Research on course management systems in higher education* (pp. 143–169). Information Age Publishing.

Moore, S. L., & Ellsworth, J. B. (2014). Ethics of educational technology. Handbook of research on educational communications and technology, 113-127.

Nakamura, W., Oliveira, E., & Conte, T. (2018). *Applying Design Science Research to develop a Technique to Evaluate the Usability and User eXperience of Learning Management Systems*. 953. https://doi.org/10.5753/cbie.sbie.2018.953

Nemoto, T., & Beglar, D. (2014). Developing Likert-Scale Questionnaires. JALT 2013 Conference Proceedings, 1–8.

Nielsen, J. (1994a). 10 Usability Heuristics for User Interface Design. Nielsen Norman Group. https://www.nngroup.com/articles/ten-usability-heuristics/

Nielsen, J. (1994b). Enhancing the explanatory power of usability heuristics. In Proceedings of the SIGCHI conference on Human Factors in Computing Systems (CHI '94), pp. 152–158.

Nielsen, J. (1995). *10 Heuristics for User Interface Design: Article by Jakob Nielsen*. Nielsen Norman Group. https://www.nngroup.com/articles/ten-usability-heuristics/

Nielsen, J., & Molich, R. (1990). Heuristic evaluation of user interfaces. *Proceedings of the SIGCHI Conference on Human Factors in Computing Systems Empowering People - CHI '90*, 249–256. https://doi.org/10.1145/97243.97281

Nokelainen, P. (2005). The technical and pedagogical usability criteria for digital learning material. In Piet Kommers & Griff Richards (Eds.), *In EdMedia+ Innovate Learning* (pp. 1011–1016). Association for the Advancement of Computing in Education (AACE). https://www.learntechlib.org/primary/p/20212/

Nokelainen, P. (2006). An empirical assessment of pedagogical usability criteria for digital learning material with elementary school students. *Journal of Educational Technology & Society*, *9*(2), 178–197.

Nunnally, J. C. (1978). *Psychometric Theory*. McGraw-Hill.

Oztekin, A., Delen, D., Turkyilmaz, A., & Zaim, S. (2013). A machine learning-based usability evaluation method for eLearning systems. In *Decision Support Systems* (Vol. 56, Issue 1, pp. 63–73). https://doi.org/10.1016/j.dss.2013.05.003

Pham, M., Xu, X., Bueno, J., & He, H. (2019). Pedagogical usability in the design of a learning module on virtual reality. *In Society for Information Technology & Teacher Education International Conference*, 547–551. https://www.learntechlib.org/primary/p/207693/

Polit, D. F., & Beck, C. T. (2006). The content validity index: Are you sure you know what's being reported? critique and recommendations. *Research in Nursing & Health*, *29*(5), 489–497. https://doi.org/10.1002/nur.20147

Preece, J. (2001). Sociability and usability in online communities: Determining and measuring success. *Behaviour & Information Technology*, *20*(5), 347–356. https://doi.org/10.1080/01449290110084683

Quiñones, D., Rusu, C., & Rusu, V. (2018). A methodology to develop usability/user experience heuristics. *Computer Standards & Interfaces*, *59*, 109–129. https://doi.org/10.1016/j.csi.2018.03.002

R Core Team (2023). _R: A Language and Environment for Statistical Computing_. R Foundation for Statistical Computing, Vienna, Austria. <https://www.R-project.org/>.

Reeves, T. C. (1994). Evaluating what really matters in computer-based education. In M. Wild & D. Kirkpatrick (Eds.), *Computer education: New perspectives* (pp. 219–246). MASTEC, Edith Cowan University.

Reeves, T. C., Benson, L., Elliott, D., Grant, M., Holschuh, D., Kim, B., Kim, H., Lauber, E., & Loh, S. (2002). Usability and Instructional Design Heuristics for E-Learning Evaluation. In P. Barker & S. Rebelsky (Eds.), *Proceedings of ED-MEDIA 2002—World Conference on Educational Multimedia, Hypermedia & Telecommunications* (pp. 1615–1621). Association for the Advancement of Computing in Education (AACE). https://www.learntechlib.org/primary/p/10234/.

Reeves, T. C., & Hedberg, J. G. (2003). *Interactive Learning Systems Evaluation*. Educational Technology.

Revelle, W. (2023). _psych: Procedures for Psychological, Psychometric, and Personality Research_. Northwestern University, Evanston, Illinois. R package version 2.3.3, <https://CRAN.R-project.org/package=psych>.

Rizopoulos, D. (2006). ltm: An R package for Latent Variable Modelling and Item Response Theory Analyses, Journal of Statistical Software, 17(5), 1-25. https://doi.org/10.18637/jss.v017.i05

Rose, D. (2000). Universal design for learning. Journal of Special Education Technology, 15(4), 47-51.

Rosseel, Y. (2012). lavaan: An R Package for Structural Equation Modeling. Journal of Statistical Software, 48(2), 1-36. https://doi.org/10.18637/jss.v048.i02

Sauro, J., & Lewis, J. (2016). *Quantifying the user experience: Practical statistics for user research* (2nd edition). Elsevier, Morgan Kaufmann.

Schmidt, M., Earnshaw, Y., Jahnke, I., & Tawfik, A. A. (2024). Entangled eclecticism: a sociotechnical-pedagogical systems theory approach to learning experience design. Educational technology research and development, 1-23.

Shi, J., Mo, X., & Sun, Z. (2012). [Content validity index in scale development]. *Zhong Nan Da Xue Xue Bao. Yi Xue Ban = Journal of Central South University. Medical Sciences*, *37*(2), 152–155. https://doi.org/10.3969/j.issn.1672-7347.2012.02.007

Shin, D. (2012). Cross‑analysis of usability and aesthetic in smart devices: What influences users' preferences? *Cross Cultural Management: An International Journal*, *19*(4), 563–587. https://doi.org/10.1108/13527601211270020

Silius, K., Tervakari, A.-M., & Pohjolainen, S. (2003). A multidisciplinary tool for the evaluation of usability, pedagogical usability, accessibility and informational quality of web-based courses. *The Eleventh International PEG Conference: Powerful ICT for Teaching and Learning*, *28*, 1–10.

Smith, P., Dugan, S., & Trompenaars, F. (1996). National Culture and the Values of Organizational EmployeesA Dimensional Analysis Across 43 Nations. *Journal of Cross-Cultural Psychology - J CROSS-CULT PSYCHOL*, *27*, 231–264. https://doi.org/10.1177/0022022196272006

Smith, P. L., & Ragan, T. J. (2004). Instructional design. John Wiley & Sons.

Squires, D., & Preece, J. (1996). Usability and learning: Evaluating the potential of educational software. *Computers & Education*, *27*(1), 15–22. https://doi.org/10.1016/0360-1315(96)00010-3

Squires, D., & Preece, J. (1999). Predicting quality in educational software: Evaluating for learning, usability and the synergy between them. *Interacting with Computers*, 17.

Tervakari, A.-M., & Pohjolainen, S. (2013). A multidisciplinary tool for the evaluation of usability, pedagogical usability, accessibility and informational quality of web-based courses. *In The Eleventh International PEG Conference: Powerful ICT for Teaching and Learning*, *28*, 11.

Tomlinson, C., Brighton, C., Hertberg, H., Callahan, C., Moon, T., Brimijoin, K., Conover, L., & Reynolds, T. (2003). Differentiating Instruction in Response to Student Readiness, Interest, and Learning Profile in Academically Diverse Classrooms: A Review of Literature. Journal for the Education of the Gifted, 27, 119–145. https://doi.org/10.1177/016235320302700203

Tselios, N., Avouris, N., & Komis, V. (2008). The effective combination of hybrid usability methods in evaluating educational applications of ICT: Issues and challenges. *Education and Information Technologies*, *13*(1), 55–76. https://doi.org/10.1007/s10639-007-9045-5

Vatrapu, R. K. (2008). CULTURAL CONSIDERATIONS IN COMPUTER SUPPORTED COLLABORATIVE LEARNING. *Research and Practice in Technology Enhanced Learning*, *03*(02), 159–201. https://doi.org/10.1142/S1793206808000501

Webb, S. M., Prieto, L., Badia, X., Albareda, M., Catalá, M., Gaztambide, S., Lucas, T., Páramo, C., Picó, A., Lucas, A., Halperin, I., Obiols, G., & Astorga, R. (2002). Acromegaly Quality of Life Questionnaire (ACROQOL) a new health-related quality of life questionnaire for patients with acromegaly: Development and psychometric properties. Clinical Endocrinology, 57(2), 251–258. https://doi.org/10.1046/j.1365-2265.2002.01597.x

Whitbeck, C. (1996). Ethics as Design: Doing Justice to Moral Problems. Hastings Center Report, 26(3), 9–16. https://doi.org/10.2307/3527925

Wolfe, E. W., & Smith Jr., E. V. (2007). Instrument development tools and activities for measure validation using Rasch models: Part I--instrument development tools. Journal of Applied Measurement, 8, 97–123.

Worthington, R. L., & Whittaker, T. A. (2006). Scale Development Research: A Content Analysis and Recommendations for Best Practices. *The Counseling Psychologist*, *34*(6), 806–838. https://doi.org/10.1177/0011000006288127

Yang, T., Linder, J., & Bolchini, D. (2012). DEEP: Design-Oriented Evaluation of Perceived Usability. *International Journal of Human–Computer Interaction*, *28*(5), 308–346. https://doi.org/10/bpngdg

Yeo, A. (1998, April). Cultural effects in usability assessment. In CHI 98 conference summary on Human factors in computing systems (pp. 74-75).

Yusoff, M. S. B. (2019). ABC of Content Validation and Content Validity Index Calculation. *Education in Medicine Journal*, *11*(2), 49–54. https://doi.org/10.21315/eimj2019.11.2.6

Zaharias, P. (2004). Usability and e-Learning: The road towards integration. *ACM ELearn Magazine*, *6*. https://doi.org/10.1145/998337.998345

Zaharias, P. (2008). Cross-Cultural Differences in Perceptions of E-Learning Usability: An Empirical Investigation. *International Journal of Technology and Human Interaction (IJTHI)*, *4*(3), 1–26. https://doi.org/10.4018/jthi.2008070101

Zaharias, P., & Poylymenakou, A. (2009). Developing a usability evaluation method for e-learning applications: Beyond functional usability. In *International Journal of Human-Computer Interaction* (Vol. 25, Issue 1, pp. 75–98).

# Appendix A

**A comparison of items in Alpha 1.1 and Alpha 1.2, including the type of changes incorporated.**

| Items in Alpha 1.1 | Items in Alpha 1.2 | Changes Made to the Original Item |
|---|---|---|
| The application was easy to use. | The application is easy to use. | Revised |
| I was able to finish tasks quickly without having to click or scroll excessively. | I am able to complete tasks quickly when using this application. | Revised |
| The application included features that made it easier to use for individuals with disabilities. | In my opinion, this application is accessible for individuals with disabilities. | Revised |
| If I came across technical errors, I was able to fix them. | This application is helpful in recovering from errors, if encountered. | Revised |
| I like how the learning application looks in general. | I like the look-and-feel of this application. | Revised |
| Moving around to different sections of this application was straightforward. | Navigating within this application is straightforward. | Revised |
| It was easy to learn how to use this application. | It is easy to learn how to use this application. | Revised |
| I was able to successfully finish tasks and scenarios with this application. | -- | Removed |
| I feel like I had control over what I wanted to do when using this application. | I feel like I have control over what I need to do when using this application. | Revised |
| -- | This application has all the functions I expect it to have for achieving my learning goals. | Newly added |
| The learning materials provided by this application were of excellent quality. | Content provided by this application is of good quality. | Revised |
| I felt fully engaged and absorbed while learning with this application. | I feel engaged while learning with this application. | Revised |
| I think this application was important and useful for reaching the learning goals. | I think this application is useful for reaching my learning goals. | Revised |
| The application gave me feedback on how well I understood the learning materials. | This application provides feedback on how well I understand the content. | Revised |
| -- | This application provides features that allow me to evaluate my performance. | Newly added |
| Interacting and collaborating with others through this application aided in my own learning. | This application provides opportunities for collaborating with others. | Revised |
| The learning materials in this application did not feel overwhelming to me. | Content in this application does not feel overwhelming to me. | Revised |
| The application provided enough support to help me reach the learning goals. | The application provides enough support in my learning process. | Revised |
| I could manage the speed and order of my own learning with this application. | I can manage the pace of my learning with this application. | Revised |
| I did not experience any prejudices based on culture, race, or gender when using the application/material. | I do not experience prejudices based on culture, race, gender, etc. when using this application. | Revised |
| In my opinion, this application could be used by learners from diverse backgrounds. | In my opinion, this application could be used by a range of learners, regardless of their social, cultural, or demographic backgrounds. | Revised |
| Overall, I am satisfied with using the application. | Overall, I am satisfied with using the application. | No change |

## Appendix B

**Items in Alpha 1.3.**

| Item # | Items in Alpha 1.3 |
|---|---|
| 1 | The application is easy to use. |
| 2 | I am able to complete tasks quickly when using this application. |
| 3 | In my opinion, this application is accessible to all individuals, regardless of their physical, cognitive, and sensory abilities. |
| 4 | This application is helpful in recovering from errors, if encountered. |
| 5 | I like the look-and-feel of this application. |
| 6 | Navigating within this application is straightforward. |
| 7 | It is easy to learn how to use this application. |
| 8 | I feel like I have control over what I need to do when using this application. |
| 9 | This application has all the functions I expect it to have for achieving my learning goals. |
| 10 | Content provided by this application is of high quality. |
| 11 | I feel engaged while learning with this application. |
| 12 | I think this application is useful for reaching my learning goals. |
| 13 | This application provides feedback on how well I understand the content. |
| 14 | This application provides features that evaluate my performance. |
| 15 | This application provides opportunities for collaborating with others. |
| 16 | Content in this application does not feel overwhelming to me. |
| 17 | This application provides enough support in my learning process. |

| 18 | I can manage the pace of my learning with this application. |
|---|---|
| 19 | I do not experience prejudices based on culture, race, gender, or other factors when using this application. |
| 20 | In my opinion, this application could be used by a range of learners, regardless of their social, cultural, or demographic backgrounds. |
| 21 | Overall, I am satisfied with this application. |

# Appendix C

**Loadings of the rotated factors with items in Alpha 1.3 and their alpha coefficient.**

|  | Factor 1 | Factor 2 | Factor 3 |
|---|---|---|---|
| **Factor 1 – Learner Satisfaction and Engagement** | $\alpha = 0.886$ | | |
| V5. I like the look-and-feel of this application. | 1.012[2] | | |
| V10. Content provided by this application is of high quality. | 0.886 | | |
| V12. I think this application is useful for reaching my learning goals. | 0.747 | | |
| V11. I feel engaged while learning with this application. | 0.633 | | |
| V21. Overall, I am satisfied with this application. | 0.535 | | |
| V9. This application has all the functions I expect it to have for achieving my learning goals. | 0.453 | | |
| V3. In my opinion, this application is accessible to all individuals, regardless of their physical, cognitive, and sensory abilities. | 0.378 | | |
| **Factor 2 – Ease-of-Use and Accessibility** | $\alpha = 0.876$ | | |
| V7. It is easy to learn how to use this application. | | 1.047[1] | |
| V6. Navigating within this application is straightforward. | | 0.905 | |
| V2. I am able to complete tasks quickly when using this application. | | 0.731 | |
| V8. I feel like I have control over what I need to do when using this application. | | 0.713 | |
| V20. In my opinion, this application could be used by a range of learners, regardless of their social, cultural, or demographic backgrounds. | | 0.661 | |
| V1. The application is easy to use. | | 0.640 | |
| V16. Content in this application does not feel overwhelming to me. | | 0.616 | |

| | | |
|---|---|---|
| V19. I do not experience prejudices based on culture, race, gender, or other factors when using this application. | | 0.410 |
| **Factor 3 – Learning Support** | α = 0.806 | |
| V14. This application provides features that evaluate my performance. | | 0.829 |
| V17. This application provides enough support in my learning process. | | 0.733 |
| V13. This application provides feedback on how well I understand the content. | | 0.683 |
| V15. This application provides opportunities for collaborating with others. | | 0.469 |
| V18. I can manage the pace of my learning with this application. | | 0.455 |
| V4. This application is helpful in recovering from errors, if encountered. | | 0.332 |

²*The loading of this variable on Factor 1 was greater than 1 because a polychoric correlation matrix was used, which was more appropriate for analyzing ordinal data than using Pearson's correlations. A pattern loading can lie beyond [-1, 1].*

# Appendix D

**Factor diagram for Beta 1.1, the final version of CAUSLT.**

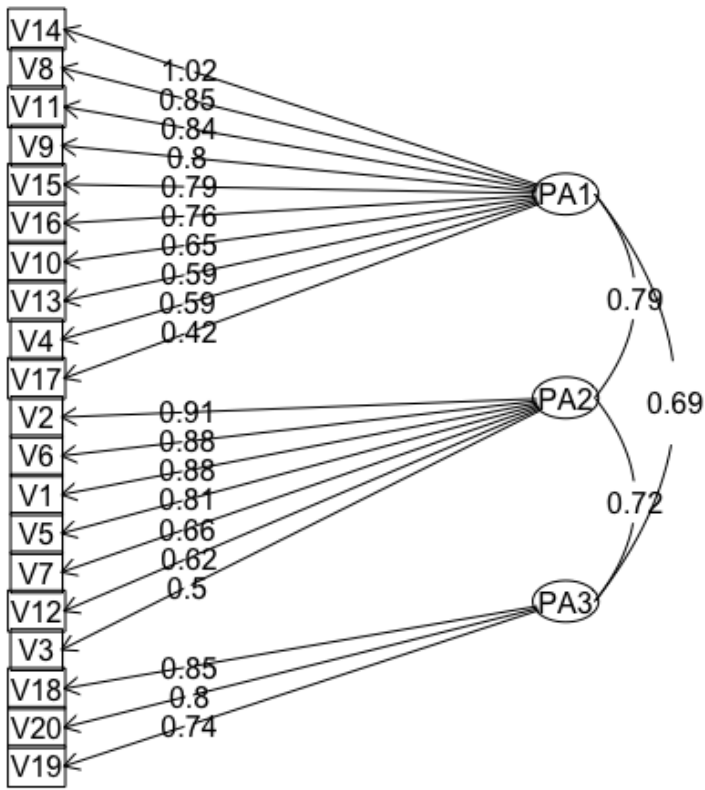